\documentclass[runningheads]{llncs}
\usepackage{graphicx}
\usepackage{multirow}
\usepackage{todonotes}
\usepackage{hyperref} 

\begin{document}
\title{Acoustic models of Brazilian Portuguese Speech based on Neural Transformers}
\titlerunning{Acoustic Models for Brazilian Portuguese Speech}
%
\author{Marcelo Matheus Gauy\inst{1},\\
Marcelo Finger\inst{1}}
\authorrunning{Gauy and Finger 2022}
%
\institute{Universidade de S\~{a}o Paulo}
\maketitle              
\begin{abstract}

An acoustic model, trained on a significant amount of unlabeled data, consists of a self-supervised learned speech representation useful for solving downstream tasks, perhaps after a fine-tuning of the model in the respective downstream task. In this work, we build an acoustic model of Brazilian Portuguese Speech through a Transformer neural network. This model was pretrained on more than $800$ hours of Brazilian Portuguese Speech, using a combination of pretraining techniques \cite{liu2020mockingjay}. Using a labeled dataset collected for the detection of respiratory insufficiency in Brazilian Portuguese speakers \cite{spira2021}, we fine-tune the pretrained Transformer neural network on the following tasks: respiratory insufficiency detection, gender recognition and age group classification. We compare the performance of pretrained Transformers on these tasks with that of Transformers without previous pretraining, noting a significant improvement.  In particular, the performance of respiratory insufficiency detection obtains the best reported results so far, indicating this kind of acoustic model as a promising tool for speech-as-biomarker approach. Moreover, the performance of gender recognition is comparable to the state of the art models in English.

\textit{Under review at Journal of the Brazilian Computer Society}
\end{abstract}
\section{Introduction}
\label{sec:intro}

This work was motivated by an Artificial Intelligence project during the COVID-19 pandemic to detect respiratory insufficiency via speech analysis \cite{SPIRA-PMLD2022,SpiraAccoustic2021}. Here we study whether it is possible to exploit unsupervised speech representation learning techniques to obtain the best possible neural model for respiratory insufficiency detection. In addition, we explore the applications of our model in other domains.

Automated speech representation learning has the goal of transforming speech signals into a format suitable and more accessible to downstream voice processing tasks. Speech signals possess a rich set of acoustic and linguistic content, containing words, phonemes, tones, semantic meanings and more. 
An acoustic model is trained on a large dataset of unlabeled data with self-supervision to acquire a general form of speech representation. This speech representation can later be used to obtain better performance on downstream tasks, where data availability may be limited.

In this paper, we study the Transformer~\footnote{Specifically, we use the Transformer encoder units from \cite{vaswani2017attention}} architecture \cite{vaswani2017attention} as a means to build an acoustic model of Brazilian Portuguese Speech. Transformers have been shown to be very effective when training is divided in two phases: \textbf{pretraining} and \textbf{refinement} \cite{devlin2018bert}. In the \textit{pretraining} phase, the Transformer is trained on large quantities of generic data using a generic task. Here, we pretrain Transformers, building an acoustic model on large quantities of Brazilian Portuguese audio data, leveraging several pretraining techniques (generic tasks) called Masked acoustic modeling \cite{liu2020mockingjay,liu2020tera}. In the \textit{refinement} phase, the pretrained acoustic model is fine-tuned on a smaller task specific dataset. Here, we use a small dataset of audio samples collected during the COVID-19 pandemic in hospitals and over the internet to enable training of a model that detects respiratory insufficiency \cite{spira2021}. The datasets are available in Zenodo (see Section~\ref{subsec:pretrainingcorpora} and Section~\ref{subsec:refinement_datasets}).
We test our model in three tasks based on this dataset: respiratory insufficiency detection (whether blood oxygen saturation is below $92\%$), gender recognition (whether male or female voice) and age group classification (whether below $40$, between $40$ and $60$ or above $60$).

We train two types of acoustic Transformers: MFCC-gram Transformers, which are fed Mel Frequency Cepstral Coefficients (MFCC) extracted from the audio files, and Spectrogram Transformers, which are fed mel spectrogram features extracted from the audio files. We use three different pretraining techniques and in addition a fourth one that combines all three. In total, this produces eight acoustic models of Brazilian Portuguese speech which will be tested on the three aforementioned tasks. We expect our pretrained models to be useful for other researchers of Brazilian Portuguese as they can be quickly fine-tuned on their specific tasks.

We compare performance of Transformers with and without pretraining. A \textit{baseline Transformer} is one where pretraining is just a random assignment of weights. We compare Transformers which undergo some form of pretraining technique with the baseline Transformer performance on all three tasks. Pretraining is performed on about $800$ hours of Brazilian Portuguese audio data using one of the four pretraining techniques aforementioned. Generally, we find that Transformers performance is improved over the baseline when using some form of pretraining on all three tasks, with the most significant differences observed in respiratory insufficiency detection and gender recognition.

As discussed in Section~\ref{sec:results}, accuracy of respiratory insufficiency detection attains the best results reported so far at $97.40\%$ (up from $96.53\%$ in \cite{gauy2021audio}), leveraging the large amount of data used in pretraining. To the best of our knowledge, no previous results in Brazilian Portuguese speech were available for gender recognition and age group classification. However, in English, there are a lot of datasets available and a few recent works \cite{kwasny2021gender,ferreira2021comparison,tursunov2021age,sanchez2022age}. For gender recognition, we attain comparable  accuracy ($98.69\%$ opposed to $99.6\%$ in  \cite{kwasny2021gender}) to the best performing work on the TIMIT dataset \cite{zue1990speech}. Note that TIMIT is English and our pretraining data was in Brazilian Portuguese (unlike the authors of \cite{kwasny2021gender} which used large English datasets for pretraining). The dataset from \cite{spira2021} is seemingly harder and we attain $93.55\%$ accuracy on it. We do not know whether the extra difficult stems from the language or other properties of the dataset~\footnote{The dataset from \cite{spira2021} is rather small and contains about one hour of audio data. However, by contacting the authors, we were able to obtain the entire dataset collected over the internet. This is not useful for respiratory insufficiency detection as the extra data comes from the control group which does not suffer from respiratory insufficiency. It is useful though for gender recognition and age group classification and we are left with a dataset containing about $18$ hours of audio data, a bit larger than TIMIT. As a result, we use the original $1$ hour dataset from \cite{spira2021} for the task of respiratory insufficiency and the complete $18$ hours dataset for the tasks of gender recognition and age group classification.}. For age group classification, in the dataset from \cite{spira2021} (i.e. for Brazilian Portuguese speech), we attain $48.69\%$ accuracy, which is marginally above what a naive majority class classifier would achieve ($45\%$). Age group classification is a substantially harder task and it should not be surprising that the accuracy is substantially lower. Moreover, in English, our models do not perform as well as the known state of the art models \cite{sanchez2022age} and it seems Transformers is not really suited for the task. Even the best performing Transformer model attains only $58.3\%$ accuracy on Common Voice (as opposed to around $80\%$ for the models from \cite{sanchez2022age}). We refer the reader to Section~\ref{subsec:comparison_english} for further discussion.


\section{Related work}
There is a long history of unsupervised speech representation learning techniques \cite{chorowski2019unsupervised,chung2018speech2vec,chung2016audio,oord2018representation,chung2019unsupervised,schneider2019wav2vec,baevski2020wav2vec,baevski2019vq,kawakami2019unsupervised,liu2019unsupervised}.
Acoustic features such as Mel Spectrograms or MFCCs are commonplace in the area of speech representation learning.  Mel spectrogram is a spectrogram that is converted to a Mel scale, which  mimics how the human ear works, scaling frequencies logarithmically so as to  emphasize lower frequencies. MFCCs or Mel Frequency Cepstral Coefficients are obtained by applying two Fourier transforms to the audio waveforms. Directly, or through the use of neural networks, they are very useful features for speech representation. 
However, even more robust features and higher level information can be extracted from speech through the use of unsupervised speech representation learning. Moreover, unlabeled data is much more abundantly available than labeled data. Thus extracting useful representation from unlabeled data is an extremely cost efficient way to improve the model quality.

Transformers were designed for natural language text processing \cite{vaswani2017attention,devlin2018bert}. More recently, they were used in audio processing tasks \cite{liu2020mockingjay,liu2020tera,schneider2019wav2vec,baevski2020wav2vec,baevski2019vq,song2019speech}. The XLNet maximizes the expected log-likelihood of a word sequence with respect to all possible autoregressive factorization orders \cite{yang2019xlnet}. In Speech-XLNet a speech based version of it is proposed \cite{song2019speech}.  In Mockingjay and Tera \cite{liu2020mockingjay,liu2020tera}, it was shown that variants of the Cloze task \cite{taylor1953cloze,devlin2018bert} for audio could be used in the pretraining phase of Transformers; this is described in Section~\ref{subsec:pretraining_techniques}. The model performance was analysed in phoneme and speaker recognition tasks. The Wav2Vec method and its variants \cite{schneider2019wav2vec,baevski2020wav2vec,baevski2019vq} use a contrastive loss to pretrain Transformers.

For respiratory insufficiency detection in Brazilian Portuguese speech, there are works prior to this one in \cite{spira2021,gauy2021audio}. 
We have found no works on respiratory insufficiency detection in English, but there are initiatives which try to detect COVID-19 from voice in \cite{pinkas2020sars,laguarta2020covid,despotovic2021detection}. 
Moreover, there was a competition (ComParE 2021) to detect COVID-19 from cough and speech data \cite{casanova21_interspeech}. 
For gender recognition and age group classification, we have not found corresponding works in Brazilian Portuguese but there are quite a few recent works for those tasks in English \cite{kwasny2021gender,ferreira2021comparison,tursunov2021age,sanchez2022age}.
Most of these attain an error rate below $2\%$ for gender recognition, but age group classification (or age estimation if they are using regression) is significantly harder and the models typically have around $20\%$ error rate or more. 
Lastly, in Brazilian Portuguese speech, there was recently the Speech and Emotion Recognition Challenge ($SE\&R$ 2022) which is a related classification task for which our pretrained models can be used \cite{gauy2022pretrained}.

\section{Methodology}

In this section, we describe the main techniques and methods used in this paper. As mentioned in Section~\ref{sec:intro}, our Transformer models are trained in two phases, pretraining and fine-tuning \cite{devlin2018bert}. In Section~\ref{subsec:pretrainingcorpora}, we describe the datasets used for pretraining the models in Brazilian Portuguese Speech. In Section~\ref{subsec:refinement_datasets}, we describe the datasets used when finetuning the model for the downstream tasks of respiratory insufficiency detection, gender recognition and age group classification. In Section~\ref{subsec:preprocessing}, we describe the general preprocessing steps that we have to undertake with the audios before feeding them as inputs to our models. In Section~\ref{subsec:Transformers}, we give a complete description of the Transformer architecture we used in our three downstream tasks. In Section~\ref{subsec:pretraining_techniques}, we describe the pretraining techniques \cite{liu2020mockingjay,liu2020tera} we used when pretraining our Transformer models.

\subsection{Pretraining Datasets}
\label{subsec:pretrainingcorpora}
For the unsupervised pretraining phase of Transformers, we employ datasets containing Brazilian Portuguese Speech. The datasets may be downloaded at \href{https://zenodo.org/record/6794924}{Zenodo}.
In this work, we use NURC-S\~{a}o Paulo \cite{castilho1986nurc}, NURC-Recife \cite{oliviera2016nurc}, ALIP \cite{gonccalves2019projeto}, C-Oral Brasil \cite{raso2012c}, SP2010 \cite{mendes2013projeto} and Programa Certas Palavras \cite{teixeira1997certaspalavras}. NURC-S\~{a}o Paulo (around $345$ hours) and NURC-Recife (around $215$ hours) are part of the NURC project which collected Brazilian Portuguese spontaneous and prepared speech in the cities of S\~{a}o Paulo and Recife. It contains dialogues between two speakers and formal utterances. ALIP (around $75$ hours) is a project that collected Brazilian Portuguese spontaneous speech from the interior of S\~{a}o Paulo. It contains interviews with people of different social backgrounds and dialogues collected in contexts of free social interaction. C-Oral Brasil (around $15$ hours) is a project that collected spontaneous speech from Minas Gerais. It contains monologues and dialogues of spontaneous nature. SP2010 (around $45$ hours) is a project containing spontaneous and read speech from S\~{a}o Paulo. It is stratified according to gender, age and scholarship level, with each level containing $5$ speakers from different regions of S\~{a}o Paulo. Programa Certas Palavras (around $125$ hours) contains collections of a radio program on air during 1981 to 1996. It contains interviews (spontaneous and prepared speech) about books and other topics. Together all these corpora contain more than $800$ hours of audio data. A smaller version of the pretraining datasets used here have been made available as the CORAA dataset \cite{junior2021coraa} for automatic speech recognition on Brazilian Portuguese. The version $1$ of the CORAA dataset contains NURC-Recife, ALIP, C-Oral Brasil and SP2010. Versions $2$ and $3$ will contain also NURC-S\~{a}o Paulo and Programa Certas Palavras.

\subsection{Refinement datasets}
\label{subsec:refinement_datasets}

In order to fine-tune Transformers for specific tasks, we need a labeled dataset containing the necessary information about the speakers. For respiratory insufficiency detection, we use the same dataset as was used in \cite{spira2021}. For gender recognition and age group classification, we increment the dataset from \cite{spira2021} with the entire control data collected over the internet (by asking the authors for the data). Both the dataset used in \cite{spira2021} and the complete $18$ hours dataset of all control data can be downloaded at~\href{https://zenodo.org/record/6672451}{Zenodo}.

The dataset from \cite{spira2021} is a relatively small (about $50$ minutes) COVID-19 dataset on respiratory insufficiency, which contains data about age and sex of each speaker. It was collected by medical students at COVID wards during the first wave of the pandemic, from patients with blood oxygen saturation level below $92\%$ as an indication of respiratory insufficiency. The control data was collected via voice donations over the internet without access to blood oxygenation levels and were assumed healthy. The control data is significantly more numerous than the patient data and for respiratory insufficiency, the authors of \cite{spira2021} take advantage of the extra control data to balance the dataset by sex. So, for each patient audio file collected, a similar control audio file is selected (same sex). For gender recognition and age group classification, we do not need to balance the training set and can use the entire original control dataset (thus, adding an extra $18$ hours to the dataset). As shown in Table~\ref{table:dataset}, the train dataset is used for finetuning in the task of respiratory insufficiency detection, whereas the train-extra dataset is used for finetuning in the tasks of gender recognition and age group classification.

Since COVID wards are extremely noisy, an extra collection was made consisting of samples of pure background noise (no voice), usually at the start of a collecting session. As there are background noise differences in data collection, the extra collection is an essential step in preventing overfitting (in the respiratory insufficiency detection task).

The gathered audio files contained $3$ utterances:
\begin{itemize}
\item Long sentence with $31$ syllables. Designed by linguists to be simple enough for low literacy donors to speak while long enough to have reading pauses.
\item Nursery rhyme for readers with reading impediments
\item Song along the lines of 'Happy birthday to you'.
\end{itemize}

Unlike other COVID-19 datasets \cite{casanova21_interspeech}, this dataset contains only speech and no cough sounds.
For simplicity, the authors of \cite{spira2021} selected only audio files of the first utterance. To deal with the presence of ward background noise, they inserted ward noise to the control group as that is better than removing it from the patients' signal. The main advantage is that this avoids inadvertently eliminating from the signal acoustic information that may be relevant to the network's classification task (e.g. noise like sounds, such as, heavy breathing).  On the other hand, the signal-noise ratio decreases.

We use the same partition in training, validation and test set as done in \cite{spira2021}. Sometimes, the voice of the doctor doing the data collection can be heard in the patient audio samples. As that would trivialize the task, audio samples where no such voice could be heard were added to the test set and to the validation set. The goal of this partition, as is usual in machine learning, is to detect training overfitting. Table~\ref{table:dataset} contains information on the dataset.

%

\begin{table*}
\caption{Filtered dataset information. Train Extra is used for training in the gender recognition and age group classification tasks. Training is used for respiratory insufficiency detection task. Note that training, validation and test sets are balanced by sex, but train extra is not ($36\%-64\%$ split favoring female). The age groups split is $35\%-45\%-20\%$ between the three groups in the test set.}
\centering
\begin{tabular*}{0.775\textwidth}{|c| c| c| c| c| c| c| c|}
\hline
\multicolumn{3}{|c|}{Sets} & Training & Train Extra & Validation & Test \\ \hline
   \multirow{6}{*}{Control} & \multirow{3}{*}{Male} & Age<40 & $25$ & $1283$ & $3$ & $10$ \\ \cline{3-7}
   & & 40<=Age<60 & $24$ & $968$ & $4$ & $8$ \\\cline{3-7}
   & & Age>=60 & $10$ & $399$ & $1$ & $4$ \\\cline{2-7}
   & \multirow{3}{*}{Female} & Age<40 & $38$ & $2029$ & $6$ & $10$ \\\cline{3-7}
   & & 40<=Age<60 & $31$ & $1921$ & $2$ & $17$ \\\cline{3-7}
   & & Age>=60 & $15$ & $864$ & $0$ & $5$ \\  \hline
   \multirow{6}{*}{Patients} & \multirow{3}{*}{Male} & Age<40 & $11$ & $11$ & $1$ & $9$ \\ \cline{3-7}
   & & 40<=Age<60 & $36$ & $36$ & $4$ & $14$ \\\cline{3-7}
   & & Age>=60 & $36$ & $36$ & $3$ & $5$ \\\cline{2-7}
   & \multirow{3}{*}{Female} & Age<40 & $7$ & $7$ & $2$ & $9$ \\\cline{3-7}
   & & 40<=Age<60 & $31$ & $31$ & $5$ & $10$ \\\cline{3-7}
   & & Age>=60 & $28$ & $28$ & $1$ & $7$ \\  \hline
\end{tabular*}\label{table:dataset}

\medskip
\end{table*}

Moreover, to be able to compare ourselves with the state of the art in gender recognition, we will use the TIMIT dataset \cite{zue1990speech}, an English dataset with about $5.5$ hours containing information about gender ($3.5$ hours for the training set, $0.5$ hours for the validation set and $1.5$ hours for the test set). TIMIT is a dataset containing $630$ speakers using eight major dialects of American English. Each speaker reads ten phonetically rich sentences. We note that it is not balanced by sex as the test dataset has $16$ male speakers and $8$ female speakers. The TIMIT dataset will be used so we can compare ourselves with \cite{kwasny2021gender} on the task of gender recognition, one of the more recent works on that task. Table~\ref{table:timit_dataset} contains information on the file distribution for each gender in the TIMIT dataset.

\begin{table*}
\caption{TIMIT dataset information. The training part of the TIMIT dataset is (randomly) split ($90\%-10\%$ ratio) into the training and validation sets. The test set has a $67\%-33\%$ split between male and female speech files.}
\centering
\begin{tabular*}{0.4\textwidth}{|c| c| c|}
\hline
Sets & Training+Validation & Test \\ \hline
Male & $3260$ & $1120$ \\ \hline
Female & $1360$ & $560$ \\\hline
\end{tabular*}\label{table:timit_dataset}

\medskip
\end{table*}

In addition, we will use the Common Voice dataset to compare ourselves with the state of the art in age group classification in English. Common Voice is a large open source dataset by the Mozilla foundation where people can donate their voice to the project online. The version we used (8.0) contains about $2800$ recorded hours of English speech (not all labeled). Among the data which is labeled with age groups, the Train set contains $980$ hours, validation and test set contain each about $4$ hours. Table~\ref{table:common_voice_dataset} contains information on the file distribution for each age group in the Common Voice dataset. Observe that the age group consisting of teens and twenties is larger than the other age groups and the elderly age group (sixties, seventies and eighties) is much smaller than the other two, so the dataset is not balanced.

\begin{table*}
\caption{Common Voice dataset information. The test set has a $62\%-34\%-4\%$ split between the different age groups.}
\centering
\begin{tabular*}{0.615\textwidth}{|c| c| c| c|}
\hline
Sets & Training & Validation & Test \\ \hline
Teens, Twenties & $303216$ & $1485$ & $1414$\\ \hline
Thirties, Forties, Fifties & $256668$ & $941$ & $782$ \\\hline
Sixties, Seventies, Eighties & $61877$ & $128$ & $79$ \\\hline
\end{tabular*}\label{table:common_voice_dataset}

\medskip
\end{table*}

\subsection{Preprocessing}
\label{subsec:preprocessing}

For pretraining, data comes from the Brazilian Portuguese speech corpora referred to in Section~\ref{subsec:pretrainingcorpora}. In these corpora and in the refinement datasets, the audio files are sampled at different sampling rates. For uniformity and for dimensionality reduction reasons, we resampled each audio at $16kHz$. Furthermore, each audio file is broken into $4$ seconds chunks, with a windowing of $1$ second steps. This serves as a kind of data augmentation as an audio with $8$ seconds becomes $5$ audio samples with $4$ seconds. Furthermore, this windowing technique also solves the problem of a potential imbalance in the audio lengths between control speakers and patients with respiratory insufficiency in the refinement dataset, thus avoiding that the network pays unnecessary attention to the audio lengths.

Audio files from both pretraining and refinement datasets are preprocessed with Torchaudio $0.9.0$. Mel spectrogram or MFCCs of the audio files were extracted using default Torchaudio parameters, with $128$ coefficients being retained. By default, Torchaudio uses a Fast Fourier Transform \cite{brigham1967fast} with a window of $400$ samples and hop length of $200$ samples. We would like to point out that the windowing should be done before the extraction of the mel spectrogram or MFCCs. These preprocessing steps are similar to the ones used in \cite{spira2021}.

As previously mentioned, for the task of respiratory insufficiency detection, we need to deal with the presence of ward noise in the refinement dataset from \cite{spira2021}. This is a serious bias source, a fact which can be seen in the experiments by \cite{spira2021}. We could try to filter this noise and eliminate it. However, this might eliminate important low-energy information from the data, which could have been useful for detecting respiratory insufficiency (e.g. noise like sounds, such as, heavy breathing). It could also create extra biases, as one would require different procedures for eliminating the noise from patients and from the control group. As suggested by \cite{spira2021}, we insert the noise present in COVID wards into all audio samples. This is a simpler solution that is likely also more effective.

The dataset from \cite{spira2021} contained $16$ samples with $1$ minute each containing just the background noise from COVID-19 wards. We add these noise samples to all the training, validation and test audio files as done in \cite{spira2021}. We vary the amount of ward noise we insert to each of the audio files. During training, one or more noise samples is injected to the audio samples. These are selected randomly from the pool of noise samples. We draw the starting point at random. Moreover, a factor to change the intensity of each noise sample is drawn. This factor is limited by a maximum amplitude value that depends on the patient audio noises. The goal of this procedure is inserting noise as similar to the pre-existing noise as possible.

\subsection{Transformers}
\label{subsec:Transformers}
The Transformers we consider in this work are of two types: MFCC-gram Transformers and Spectrogram Transformers. They are equivalent except in the audio features that they receive: MFCC-gram Transformers receive the MFCCs of the audio files and Spectrogram Transformers receive mel spectrogram audio features. We test both types of Transformers in all the tasks we consider. The Transformers studied here are the Transformer encoder units from \cite{vaswani2017attention}. More precisely, we use a three layer Transformer encoder with multi-head self -attention. Each encoder layer is composed of two sub-layers. The first is a multi-head self-attention network and the second is a fully connected feed-forward layer. Each sub-layer has a residual connection followed by layer normalization \cite{ba2016layer}. The encoder layers and sub-layers produce outputs of dimension $512$. The fully connected feed forward network within each encoder layer has an inner dimension of $2048$.

To generate the sequence of tokens that is fed to the Transformers, we split the MFCC and/or Spectrogram into its frames. To each frame of the MFCC or spectrogram corresponds one token fed to the sequence. One could also consider merging multiple frames into one token but in our preliminary experiments this generally led to worse results. As the Transformers needs an intrinsic way of dealing with the position of the tokens, we use sinusoidal positional encoding, as suggested by previous works \cite{vaswani2017attention,liu2020mockingjay,pham2019very}. The input frames are projected linearly to the hidden dimension of $512$, as a direct addition of the acoustic features to positional encoding may lead to potential training failure \cite{sperber2018self,liu2020mockingjay}.

Transformers typically undergo two training phases: pretraining and refinement. In the pretraining phase, we make use of the pretraining techniques described in Section~\ref{subsec:pretraining_techniques} to build acoustic models over generic audio data. In the refinement phase, the model is refined using labeled data of the refinement datasets for each task. In some experiments, we bypass the pretraining phase by initializing the Transformers with a random assignment of weights and skipping directly to the refinement phase over the labeled dataset. This gives us a baseline performance and we call these Transformers the baseline Transformers. We consider baseline MFCC-gram Transformers and baseline Spectrogram Transformers.

The code we wrote is based on the excellent guide ``The annotated Transformer''\footnote{\url{http://nlp.seas.harvard.edu/2018/04/03/attention.html}}. The Transformers we built are smaller than the ones used in BERT \cite{devlin2018bert}, as the size of the refinement dataset from \cite{spira2021}, especially for the task of respiratory insufficiency detection, is also relatively small. Therefore, we do not expect that larger Transformers would yield significantly improved results on the tasks we trained. For larger datasets, there may be significant benefit in using larger Transformers.

\subsection{Acoustic model construction: pretraining techniques}
\label{subsec:pretraining_techniques}

Here, we present three pretraining techniques for self-supervised learning of acoustic models with Transformers. These techniques are mostly based on Masked acoustic modeling from \cite{liu2020mockingjay,liu2020tera}. This is a technique that erases a fraction of the input and reconstructs the erased parts from the remaining ones. It is akin to masked language modelling from \cite{devlin2018bert}. Observe that these are bidirectional methods and that the reconstruction depends on both the left and right context.

\textbf{Time alteration}: This technique is also called Masked acoustic modelling \cite{liu2020mockingjay}. Select frames uniformly at random up to $15\%$ of the input. More precisely, select frame chunks of a certain size\footnote{In the experiments the chunk size was $7$.} uniformly at random so that the total number of masked frames results in $15\%$. Then do one of the following for each chunk: 1) mask it to zero with probability $80\%$; 2) replace it with a random frame chunk $10\%$ of the time or 3) leave the frame chunk as it is in the remaining $10\%$ of the time. The goal of these three steps is to alleviate the mismatch between training and inference (as is also done with masked language modelling from \cite{devlin2018bert}). 

\textbf{Channel alteration}: this technique is from \cite{liu2020tera}. The goal is to randomly mask a block of consecutive quefrency (if using MFCC-grams \cite{oppenheim2004frequency}), or frequency (if using Mel spectrogram) channels to zero for all time steps of the input sequence. First, select the width $W_C$ of the block uniformly at random from $\{0,1,\ldots, W\}$, where $W$ is a $10\%$ fraction of the total number of channels. Second, sample uniformly at random a channel index $I_C$ from $\{0,1\ldots,H-W_C-1\}$,  where $H$ is the total number of channels in the input. Then, channels from $I_C$ to $I_C+W_C-1$ are masked to zero. Note that a fraction of the time none of the channels will be masked which addresses the mismatch between training and inference.

\textbf{Noise alteration}: this technique is from \cite{liu2020tera}. With probability of $10\%$, sample a random magnitude matrix with the same size as the input. Each element in the matrix is sampled from a normal distribution with zero mean and $0.2$ variance. Add the sampled Gaussian noise to change the magnitude of the real input frames. This is an extension of Masked acoustic modelling where we do not mask frames to zero but to a random value. As a fraction of the time the frames are not masked, the mismatch between training and inference is addressed.

\textbf{Time+Channel+Noise}: this technique combines all three previous techniques by applying all of them individually to the real input frames. 

\section{Results and Discussion}
\label{sec:results}

\subsection{Brazilian Portuguese Speech}

In order to test the performance of Transformers in the three tasks of respiratory insufficiency detection, gender recognition and age group classification in Brazilian Portuguese Speech, we perform two experiments. The first experiment tests the performance of Baseline Transformers on the three tasks at hand. In other words, we do not do unsupervised pretraining for the first experiment, finetuning (MFCC-gram and Spectrogram) Transformers directly on the refinement dataset from \cite{spira2021} as well as its extended version (train extra files). The goal of the first experiment is to define a baseline performance which we will try to beat through unsupervised pretraining. For the second experiment, we aim to compare the different pretraining techniques with the baseline performance. MFCC-gram Transformers and Spectrogram Transformers are pretrained with all $4$ techniques described in Section~\ref{subsec:pretraining_techniques} on more than $800$ hours of Brazilian Portuguese speech data (from Section~\ref{subsec:pretrainingcorpora}). After pretraining is done, the constructed acoustic models  are fine-tuned on the refinement dataset from \cite{spira2021} on the three tasks considered here, with gender recognition and age group classification also being finetuned on the train extra files.

For the first experiment, we bypass the pretraining phase by initializing the Transformers with random weights and training it on the three tasks on the refinement dataset. We perform each experiment for $20$ epochs and repeat the experiments $10$ times. The batch sizes are set to $16$. The results are shown in Table~\ref{table:tasks_baseline}. Observe that baseline MFCC-gram Transformers outperforms baseline Spectrogram Transformers on all three tasks.

\begin{table*}
\caption{Baseline Transformers performance on respiratory insufficiency detection, gender recognition and age group classification.} 
\centering
\begin{tabular*}{0.865\textwidth}{|c| c| c|}
\hline
Task & Model & Accuracy \\ \hline

\multirow{2}{*}{Respiratory insufficiency}   &  baseline MFCC-gram Transformers        &        $90.25\pm 3.09$     \\
\cline{2-3}   
&  baseline Spectrogram Transformers        &       $74.93\pm 2.99$     \\
\hline
\multirow{2}{*}{Gender}   &  baseline MFCC-gram Transformers     &       $87.81\pm 2.98$     \\
\cline{2-3}   
&  baseline Spectrogram Transformers        &       $87.04\pm 1.18$       \\
\hline
\multirow{2}{*}{Age group}   &  baseline MFCC-gram Transformers       &       $47.70\pm 2.00$       \\
\cline{2-3}   
&  baseline Spectrogram Transformers          &       $44.54\pm 1.88$           \\
\hline
\end{tabular*}\label{table:tasks_baseline}
\end{table*}

For the second experiment, we pretrain $4$ MFCC-gram Transformers and $4$ Spectrogram Transformers (one for each pretraining technique), for a total of $8$ acoustic models. 
Pretraining is done over the entire corpora described in Section~\ref{subsec:pretrainingcorpora}, for a total of $3$ epochs. 
Once pretraining is finished, we refine the acoustic models on the refinement dataset to check their performances on all three tasks. 
Refinement is done for $20$ epochs on the refinement dataset from \cite{spira2021}. 
Each refinement is repeated $10$ times, so the results are averaged. 
In general, performance improves once a pretraining technique is used for all three tasks. 
We attain $97.40\%$ on respiratory insufficiency detection, $93.55\%$ on gender recognition and $48.69\%$ on age group classification. 
In the table, we reported the standard deviation of the $10$ repetitions for each of the $30$ experiments.
Observe that the standard deviation of the mean is the standard deviation of the repetitions divided by the root of the number of repetitions performed (therefore, divided by $\sqrt{10}$).
This implies that the standard deviation of the mean is typically between $0.5\%$ and $1\%$ for each experiment and sometimes being even lower than that. From that, one can safely conclude that in most cases, specially for respiratory insufficiency detection and gender recognition, the differences in performance cannot be explained solely by statistical fluctuations.
As a result, we can state that the performance does typically improve once a pretraining technique is used, especially for respiratory insufficiency detection and gender recognition.
We emphasize that the performance on respiratory insufficiency detection is the highest attained. As the dataset from \cite{spira2021} contains different procedures to obtain the patient and control data, one could wonder whether the performance does not come from spurious factors beyond what one is able to control by inserting the ward noise.
While this is a valid criticism, evidence we have at the moment indicates that this is likely not the case. 
In \cite{fernandessvartman22_speechprosody}, the authors propose a naive classifier which split the audios in a certain number of bins, checking which bin corresponds to moments of pause and speech. The naive classifier estimates the probability of a certain audio coming from someone suffering respiratory insufficiency as the product of the probabilities of each specific bin being a pause given that it is from a person suffering respiratory insufficiency. 
In a similar manner, one can determine the probability that a certain audio came from someone not suffering from respiratory insufficiency. The naive classifier selects the larger of the two probabilities and classifies the audio according to that. 
This naive classifier, which considers only pauses and interprets each bin independently from other bins, attains around $88\%$ accuracy. This indicates that substantial performances can be attained by even simple models and that a Transformers model may surpass that and attain really high accuracy on the respiratory insufficiency detection task. Moreover, the authors of \cite{spirainterpretability2022} investigate deep neural networks performance, akin to our Transformers, through attention maps indicating where these networks are paying attention to in the audio. 
They use the same dataset from \cite{spira2021} and their networks attain performance a little bit higher than what is reported in \cite{spira2021} though a little lower than here. Their analysis on the attention maps of their networks indicates that it concentrates on high energy areas of the audio, and in particular, positions of the audio containing speech. 
Moreover, it also seems to concentrate on moments of speech around where pauses are expected. Both results indicate that deep networks trained on the dataset from \cite{spira2021} can attain high performance and not rely on spurious information pertaining to the differences in data collection procedures on the dataset \cite{spira2021}. While the performance of gender recognition is lower than what is commonly reported on the literature \cite{kwasny2021gender}, it seems that this is because the dataset used is somehow harder than TIMIT, which is commonly reported on other works. 
This will be seen in Section~\ref{subsec:comparison_english}. Lastly, it seems age group classification is a really hard task in general and, in particular on this dataset, and results are only slightly above what one would get from predicting the majority class. 
However, even when finetuning on the large Common Voice dataset, the results obtained do not beat a naive majority class classifier and are a far cry from the state of the art models for the task \cite{sanchez2022age}. For age group classification, it seems our Transformers models are not really suited for the task and suffer to obtain good performance. 
These results will also be presented in Section~\ref{subsec:comparison_english}.

\begin{table*}
\caption{This table compares the performance of all pretraining techniques described in Section~\ref{subsec:pretraining_techniques}. The acoustic models are pretrained on Brazilian Portuguese speech data and fine-tuned on the refinement dataset from \cite{spira2021} for all three tasks. Gender recognition and age group classification are trained on the train extra files from \cite{spira2021}.} 
\centering
\begin{tabular*}{0.934\textwidth}{|c| c| c| c|}
\hline
Task & Model & Pretraining technique & Accuracy \\ 
\hline
\multirow{10}{*}{\parbox{2.1cm}{Respiratory insufficiency}}   & \multirow{5}{*}{MFCC-gram Transformers}  &   Baseline    &        $90.25\pm 3.09$ \\
\cline{3-4}
  &   &   Time alteration  &        $\mathbf{97.40\pm 1.04}$ \\
\cline{3-4}
  &     &   Channel alteration  &        $95.93\pm 1.97$ \\
\cline{3-4}
  &   &   Noise alteration  &        $96.27\pm 1.83$ \\
\cline{3-4}
  &   &   Time+Channel+Noise  &        $96.03\pm 1.88$ \\
\cline{2-4}
 & \multirow{5}{*}{Spectrogram Transformers}   &   Baseline    &       $74.93\pm 2.99$ \\
\cline{3-4}
  &   &   Time alteration  &        $78.73\pm 2.75$ \\
\cline{3-4}
  &   &   Channel alteration  &        $79.59\pm 2.72$ \\
\cline{3-4}
  &   &   Noise alteration  &        $78.05\pm 1.90$ \\
\cline{3-4}
  &   &   Time+Channel+Noise  &        $81.38\pm 2.27$ \\
\hline
\multirow{10}{*}{Gender}   & \multirow{5}{*}{MFCC-gram Transformers} &   Baseline    &        $87.81\pm 2.98$ \\
\cline{3-4}
  &   &   Time alteration  &        $89.85\pm 1.28$ \\
\cline{3-4}
  &   &   Channel alteration  &        $88.90\pm 1.82$ \\
\cline{3-4}
  &   &   Noise alteration  &        $89.68\pm 2.00$ \\
\cline{3-4}
  &   &   Time+Channel+Noise  &        $87.88\pm 2.03$ \\
\cline{2-4}
 & \multirow{5}{*}{Spectrogram Transformers}  &   Baseline    &       $87.04\pm 1.18$ \\
\cline{3-4}
  &   &   Time alteration  &        $86.39\pm 1.01$ \\
\cline{3-4}
  &   &   Channel alteration  &        $92.05\pm 2.01$ \\
\cline{3-4}
  &   &   Noise alteration  &        $86.11\pm 1.08$ \\
\cline{3-4}
  &   &   Time+Channel+Noise  &        $\mathbf{93.55\pm 0.45}$ \\
\hline
\multirow{10}{*}{Age group}   &  \multirow{5}{*}{MFCC-gram Transformers}  &   Baseline    &        $47.70\pm 2.00$ \\
\cline{3-4}
  &   &   Time alteration  &        $\mathbf{48.69\pm 2.92}$ \\
\cline{3-4}
  & MFCC-gram Transformers  &   Channel alteration  &        $48.17\pm 2.25$ \\
\cline{3-4}
  &   &   Noise alteration  &        $47.62\pm 1.23$ \\
\cline{3-4}
  &   &   Time+Channel+Noise  &        $46.91\pm 3.11$ \\
\cline{2-4}
 & \multirow{5}{*}{Spectrogram Transformers}  &   Baseline    &       $44.54\pm 1.88$ \\
\cline{3-4}
  &   &   Time alteration  &        $43.78\pm 0.85$ \\
\cline{3-4}
  &   &   Channel alteration  &        $44.30\pm 1.87$ \\
\cline{3-4}
  &   &   Noise alteration  &        $43.29\pm 1.72$ \\
\cline{3-4}
  &   &   Time+Channel+Noise  &        $43.58\pm 1.60$ \\
\hline
\end{tabular*}\label{table:comparison}
\end{table*}

\subsection{Comparison with other works for gender recognition and age group classification}
\label{subsec:comparison_english}

As we mentioned before, to the best of our knowledge, our work is the first on gender recognition in Brazilian Portuguese. In English, one of the most recent works on gender recognition is \cite{kwasny2021gender}. There, the state of the art results on the TIMIT dataset are obtained of around $99.6\%$ accuracy on gender recognition. We compare ourselves with their result, to know where we stand in relation to the state of the art. For that, we consider our $8$ pretrained models on Brazilian Portuguese speech, as well as the baseline MFCC-gram Transformers and baseline Spectrogram Transformers, and train them on the TIMIT training~\footnote{We split the training set from TIMIT randomly, using a $90\%-10\%$ ratio, into training and validation sets} set for $20$ epochs. Each time, the model performances are measured in the validation set after each epoch, and the best performing model is saved. We use batch size of $16$ and repeat each experiment $10$ times. The results are shown in Table~\ref{table:timit_comparison}. There, it can be seen that we attain $98.69\%$ accuracy which seems in line with the results of \cite{kwasny2021gender}, in particular, because we do not pretrain on datasets containing the English language. Despite pretraining only on the Brazilian Portuguese language, we still observe that the pretrained models did typically better than their respective baseline counterparts, though this difference is less pronounced than the one for the Brazilian Portuguese refinement dataset.

\begin{table*}
\caption{This table compares the performance of all pretraining techniques described in Section~\ref{subsec:pretraining_techniques}. The acoustic models are pretrained on Brazilian Portuguese speech data and fine-tuned on the TIMIT dataset for gender recognition.} 
\centering
\begin{tabular*}{0.935\textwidth}{|c| c| c| c|}
\hline
Task & Model & Pretraining technique & Accuracy \\ 
\hline
\multirow{10}{*}{TIMIT Gender}   & \multirow{5}{*}{MFCC-gram Transformers} &   Baseline    &        $98.49\pm 0.46$ \\
\cline{3-4}
  &   &   Time alteration  &        $98.57\pm 0.34$ \\
\cline{3-4}
  &   &   Channel alteration  &        $\mathbf{98.69\pm 0.23}$ \\
\cline{3-4}
  &   &   Noise alteration  &        $98.47\pm 0.32$ \\
\cline{3-4}
  &   &   Time+Channel+Noise  &        $98.35\pm 0.40$ \\
\cline{2-4}
 & \multirow{5}{*}{Spectrogram Transformers}  &   Baseline    &       $95.73\pm 1.00$ \\
\cline{3-4}
  &   &   Time alteration  &        $96.20\pm 0.66$ \\
\cline{3-4}
  &   &   Channel alteration  &        $96.36\pm 0.64$ \\
\cline{3-4}
  &   &   Noise alteration  &        $96.52\pm 0.81$ \\
\cline{3-4}
  &   &   Time+Channel+Noise  &        $96.10\pm 1.02$ \\
\hline
\end{tabular*}\label{table:timit_comparison}
\end{table*}

Lastly, we compare the performance of our work on age group classification with previous works. In Brazilian Portuguese speech, to the best of our knowledge, we are the first to study this problem. Still, in English, there are a few such works and one of the most recent and the best performing one we found is \cite{sanchez2022age}. There, the authors study age group classification in the Common Voice dataset and analyze several different models attaining around $80\%$ accuracy on the classification task. They split the labeled ages into $3$ groups: $1$) teens, twenties and $2$) thirties, fourties, fifties and $3$) sixties, seventies, eighties. Moreover, the reported accuracy is weighted average across the different groups (weights $0.25$ for the first group, $0.2$ for the second group and $0.55$ for the third group). The weighting is chosen so as to counteract the different number of elements in the different classes. To perform a proper comparison, we have trained our Transformers models on the Common Voice dataset. We selected version $8.0$ as the authors of \cite{sanchez2022age} did not mention which version they selected. As mentioned in Section~\ref{subsec:refinement_datasets}, the training set contains almost $1000$ hours of labeled data~\footnote{According to the authors of \cite{sanchez2022age} the dataset they used contained about $100$ hours of labeled data}. We did not bother to weight our accuracy with the proper weights, as the unweighted average accuracy was of only $58.3\%$, for the best performing model. This is a little bit lower than what a naive majority class classifier would achieve ($62\%$) and substantially lower than the state of the art results from \cite{sanchez2022age}. We used original training, validation and test set splits provided by Common Voice version $8.0$, restricted, of course, to the part which was labeled. We trained our models for a total of $2$ epochs (almost $250k$ steps) and saved the best performing model on the validation set (checked at the end of each epoch). The batch size used was $16$ and due to the large Common Voice dataset size we did not repeat the experiments, running each model only once (hence the lack of an estimate for sample standard deviation). The results can be seen in Table~\ref{table:common_voice_comparison}. Note that the version of Common Voice we used ($8.0$) contains almost $1000$ hours of age labeled data, so we do not believe that our Transformers failure comes from the dataset lacking enough data. The most sensible conclusion is that the Transformers we built are not really suited for the age group classification task. It is likely that better results can be attained by swapping the input modules for something other than Mel Spectrograms and MFCC-grams (perhaps, better results can be obtained by using raw audio data). We do not pursue that direction in this paper, however.

\begin{table*}
\caption{This table compares the performance of all pretraining techniques described in Section~\ref{subsec:pretraining_techniques}. The acoustic models are pretrained on Brazilian Portuguese speech data and fine-tuned on the Common Voice dataset for age group classification.} 
\centering
\begin{tabular*}{0.935\textwidth}{|c| c| c| c|}
\hline
Task & Model & Pretraining technique & Accuracy \\ 
\hline
\multirow{10}{*}{Common Voice age}   & \multirow{5}{*}{MFCC-gram Transformers} &   Baseline    &        $54.68$ \\
\cline{3-4}
  &   &   Time alteration  &        $58.05$ \\
\cline{3-4}
  &   &   Channel alteration  &        $53.92$ \\
\cline{3-4}
  &   &   Noise alteration  &        $56.3$ \\
\cline{3-4}
  &   &   Time+Channel+Noise  &        $\mathbf{58.3}$ \\
\cline{2-4}
 & \multirow{5}{*}{Spectrogram Transformers}  &   Baseline    &       $56.27$ \\
\cline{3-4}
  &   &   Time alteration  &        $53.73$ \\
\cline{3-4}
  &   &   Channel alteration  &        $55.9$ \\
\cline{3-4}
  &   &   Noise alteration  &        $53.63$ \\
\cline{3-4}
  &   &   Time+Channel+Noise  &        $54.84$ \\
\hline
\end{tabular*}\label{table:common_voice_comparison}
\end{table*}

\section{Conclusions and Future work}

We built acoustic models of Brazilian Portuguese speech by using previously known pretraining techniques. For the pretraining phase, we use data from $6$ corpora, totalling more than $800$ hours of audio. For the refinement phase, we use a dataset collected for detecting respiratory insufficiency from COVID-19. We refine the model in this dataset for: respiratory insufficiency detection, gender recognition and age group classification. We find that pretraining via different techniques typically leads to improved performance for Transformers as opposed to bypassing the pretraining phase by initializing the weights randomly.

In respiratory insufficiency detection, we attain $97.40\%$ accuracy in the dataset from \cite{spira2021} which is up from the $96.53\%$ in \cite{gauy2021audio} the previous best result reported in the literature. In gender recognition, we attain $93.55\%$ accuracy in the test set of \cite{spira2021}. To the best of our knowledge, this is the only result on Brazilian Portuguese gender recognition from voice. For gender recognition in English, more specifically in the TIMIT dataset, we attain $98.69\%$ which is comparable to the most recent result on the TIMIT dataset for the task($99.6\%$ in \cite{kwasny2021gender}). Age group classification is a significantly harder task, and we attain on the test set of \cite{spira2021} an accuracy of $48.69\%$, slightly above what one gets by predicting the majority class for that test set. Again, to the best of our knowledge, these are the first results for Brazilian portuguese speech in the task of age group classification. In English, in the task of age group classification, we attain only $58.3\%$ accuracy in the Common Voice dataset. This is actually below what a naive majority class classifier would achieve and substantially lower than what is reported in the literature \cite{sanchez2022age} (around $80\%$ accuracy). From the results for age group classification, it seems our Transformers are not really suited for the task.

Future work could involve increasing and improving the dataset of \cite{spira2021}. Ideally, the control and patient data for the new dataset would be collected in the same environment, so adding noise to the data is no longer necessary. Moreover, information  regarding the cause of respiratory insufficiency could be added to the labels, so that one can investigate whether it is possible to predict the causes of respiratory insufficiency from speech. Regarding age group classification, one could check whether changing the inputs given to our Transformers (perhaps, to raw audio data) could produce better results, as it currently seems like the Mel Spectrogram and MFCC-gram are not really suited for the task.

\section{Acknowledgements}
We would like to thank LNCC for providing the computing power required to do this work. All experiments were run in the Santos Dumont supercomputer of the LNCC servers.

This work was supported by FAPESP grant number 2020/16543-7 (POSDOC) and project 06443-5 (SPIRA).  MF was partly supported by CNPq grant PQ 303609/2018-4, Fapesp 2014/12236-1 (Animals) and  the Center for Artificial Intelligence (C4AI-USP), with support by the São Paulo Research Foundation (FAPESP grant \#2019/07665-4) and by the IBM Corporation.  This work was financed in part by the Coordenação de Aperfeiçoamento de Pessoal de Nível Superior -- Brasil (CAPES) -- Finance Code 001.

%
%
%
\bibliographystyle{acm}
\bibliography{refs}
%




\end{document}